\documentclass[preprint,prd,amsmath,amssymb,nofootinbib,tightenlines,floatfix,letterpaper]{revtex4}
\usepackage[dvips]{graphics}
\usepackage{epsfig}
\usepackage{latexsym}
\usepackage{slashed}
\newcommand{\be}{\begin{equation}}
\newcommand{\ee}{\end{equation}}
\newcommand{\bea}{\begin{eqnarray}}
\newcommand{\eea}{\end{eqnarray}}

\begin{document}
%\preprint{BNL-HET-09/}

\title{Hidden MeV-Scale Dark Matter in Neutrino Detectors}

\author{Jennifer Kile\footnote{
	Electronic address: jenkile@quark.phy.bnl.gov}
}
\affiliation{Brookhaven National Laboratory, Upton, NY 11973}

\author{Amarjit Soni\footnote{
	Electronic address: soni@bnl.gov}}
\affiliation{Brookhaven National Laboratory, Upton, NY 11973}

\begin{abstract}
The possibility of direct detection of light fermionic dark matter in neutrino detectors is explored from a model-independent standpoint.  We consider all operators of dimension six or lower which can contribute to the interaction $\bar{f} p \rightarrow e^+ n$, where $f$ is a dark Majorana or Dirac fermion.  Constraints on these operators are then obtained from the $f$ lifetime and its decays which produce visible $\gamma$ rays or electrons.  We find one operator which would allow $\bar{f} p \rightarrow e^+ n$ at interesting rates in neutrino detectors, as long as $m_f \lesssim m_{\pi}$.  The existing constraints on light dark matter from relic density arguments, supernova cooling rates, and big-bang nucleosynthesis are then reviewed.  We calculate the cross-section for $\bar{f} p \rightarrow e^+ n$ in neutrino detectors implied by this operator, and find that Super-K can probe the new physics scale $\Lambda$ for this interaction up to ${\cal O}(100 \mbox{ TeV})$.
\end{abstract}

\maketitle

\section{Introduction}
\label{sec:intro}

Despite years of effort, the identity of dark matter (DM) is still unknown.  For a review of the evidence for DM, possible DM candidates, and detection techniques, see \cite{Bertone:2004pz}.  Here, we will be concerned with direct DM detection.  Most direct searches for DM consider the case where a DM particle scatters elastically off of Standard Model (SM) particles.  As DM is nonrelativistic, $v/c \sim 10^{-3}$, the momentum transferred in these interactions is small;  for a DM particle with mass $m_{DM}\sim 100$ GeV scattering off a $100$ GeV nucleus, the momentum transfer is of order $100$ MeV.  Here, we point out that a similar momentum transfer could occur for much lighter DM particles, if they annihilate with SM particles into final-state particles with a lower rest mass.  Thus, it is reasonable to suspect that rather different mass scales could be probed with the same experiments.  For illustration, we take the interaction
\begin{equation}
f N \rightarrow f'  N'
\label{eq:proc}
\end{equation}  
where $N$ and $N'$ are SM particles (such as nuclei or nucleons), $f$ is a DM particle (for this paper, taken to be fermionic), and $f'$ is some particle which could be either contained in the SM or from new physics (NP), but with $m_{f'}<<m_f$.  In this case, the kinetic energy of the initial-state $f$ is negligible, but the rest mass $m_f$ is converted into kinetic energy of the final-state products.  If we take $m_f$ to be $O(100 \mbox{ MeV})$ and $M_N\sim M_{N'}>>m_f$, the momentum transfer can be similar to that of the more standard case given above.

This thus begs the question of whether or not current experiments could be used to rule out DM particles which are very light, but which interact with SM particles via (\ref{eq:proc}).  Recently, there has been much interest in the observability of models with ``hidden'' sectors \cite{Strassler:2006im,Patt:2006fw,Han:2007ae,ArkaniHamed:2008qn}, where new particles exist with masses similar to those of the SM fields, but whose interactions with SM particles are mediated by very heavy particles and, thus, strongly suppressed.  In this work, we consider a light (here, taken to be below the weak scale) DM particle $f$ interacting via the process (\ref{eq:proc}) for the case where $f$ is fermionic.  We take $f$ to belong to a ``hidden'' sector, i.e., its interactions with SM particles are suppressed by either a high mass scale $\Lambda$ or by a very small coupling.  Although one could consider the case where $f'$ is a hidden-sector particle or a neutrino, for simplicity, we consider only dark matter that interacts with SM particles through charged-current interactions, i.e., $f'$ is a charged, visible SM particle; specifically, for this work, we will assume that $f'$ is an electron.  

Anticipating results which will see in Section \ref{sec:dm}, the least-constrained range of DM mass is $m_f \lesssim O(100 \mbox{ MeV})$.  Although we will review the current constraints on MeV-scale DM in section \ref{sec:dm}, we point out here that DM in this mass range can have an acceptable lifetime and relic density and remain compatible with a wide variety of observational constraints.  The possible range for $m_f$ raises the possibility that the processes $f n \rightarrow p e^-$ and $\bar{f} p \rightarrow e^+ n$ could be seen in present-day neutrino detectors which search for neutrinos with energies of $10$'s of MeV.  It is the observability of $\bar{f} p \rightarrow e^+ n$ which we explore in this paper; we will specifically examine constraints obtainable from the search for relic neutrinos from supernovae performed by Super-Kamiokande \cite{Malek:2002ns}.

In this work, we consider the process $\bar{f} p \rightarrow e^+ n$ from a model-independent standpoint via effective operators.  We do this for two reasons.  First, we choose the process $\bar{f} p \rightarrow e^+ n$ based on its experimental observability; as its experimental signature is very similar to SM neutrino interactions, this process can be studied with present-day neutrino experiments.  Second, little is known of the identity (or identities) of dark matter or of possible interactions within the ``dark sector'' or between dark matter and SM particles; these interactions may turn out to be highly complicated and not well-described by any current model.  Given these considerations, we feel that any plausible method of direct detection of dark matter should be investigated, and that a model-independent analysis is in order.  

We point out that our work is somewhat similar in spirit to works on inelastic DM \cite{TuckerSmith:2001hy,Bernabei:2002qr,TuckerSmith:2002af,TuckerSmith:2004jv,Chang:2008gd,Cui:2009xq,Alves:2009nf}, but with a few important differences.   First, the range of DM mass which we consider is very different, $O(100 \mbox{ MeV})$ instead of $O(100 \mbox{ GeV})$.  Second, we consider DM which scatters to a lower-mass state; in our case, this lower-mass state is an SM particle.\footnote{DM scattering to a lower-mass state has been considered previously; see \cite{Finkbeiner:2009mi,Batell:2009vb}; for inelastic, low-mass DM in DAMA/NaI, see \cite{Bernabei:2008mv}.}  Lastly, we point out that in this work, we need not assume that elastic interactions such as $\bar{f}p\rightarrow \bar{f}p$ are suppressed; they could occur, but would have momentum transfers of $O(100 \mbox{ keV})$, and thus typical proton recoil energies of $O(10 \mbox{ eV})$, below the energy threshold of neutrino experiments.  Although the inelastic interaction $\bar{f}p\rightarrow \bar{\nu} p$ could occur with a sizeable momentum transfer, the operators which we study in section \ref{sec:dm} which contribute to this process turn out to be severely constrained.\footnote{Here, we do not include operators which contain both $f$ and non-SM right-handed neutrino fields.}

This paper is arranged as follows.  In section \ref{sec:ops}, we give the basis of operators which can contribute to the process $\bar{f} p \rightarrow e^+ n$.  We derive constraints from DM lifetime in \ref{sec:dm}, where we find that most of the operators are tightly constrained, but that the process $\bar{f} p \rightarrow e^+ n$ can still occur if the DM mass is $\lesssim O(100 \mbox{ MeV})$.  Given this result, we then review the existing constraints on MeV DM.   In section \ref{sec:nus}, we then give the scattering cross-section and give limits on the operator contributing to this process from the Super-K relic supernova neutrino search.  For the majority of this work, we consider only the case where the SM lepton and quark fields are from the first generation.  However, in section \ref{sec:collflav}, we discuss how our results change if we consider the same operators, but with different flavors for the quark and lepton fields; we also briefly mention issues involved in relating our results to specific models.  Finally, in section \ref{sec:conc}, we conclude.
\section{Operator Basis}
\label{sec:ops}
In this section, we enumerate our operator basis.  First, we summarize our notation.  We denote the lepton and quark doublets of the SM by $L$ and $Q$, respectively.  The corresponding SM right-handed singlet fields will be denoted as $\ell_R$, $u_R$, and $d_R$.  We denote the Higgs doublet as $\phi$, and define $\tilde{\phi}= i \tau^2 \phi^*$.  For now, we take all quark and lepton fields to be from the first generation; we delay discussion of other quark and lepton flavors to section \ref{sec:collflav}.

For simplicity, we consider only fermionic (Majorana or Dirac) DM particles, which we will denote as $f$.  We assume that $f$ is a singlet under the SM gauge group, $SU(3)\times SU(2)\times U(1)$.\footnote{For other studies of fermionic singlet DM, see \cite{Kim:2008pp,Kim:2009ke,Lee:2008xy,Babu:2007sm}.}  We consider operators of dimension six or less which are $SU(3)\times SU(2)\times U(1)$-invariant and can contribute to the process $f N \rightarrow f'  N'$ where $N$, $N'$, and $f'$ are all SM particles.  Additionally, we require that $N$, $N'$, and $f'$ are {\it visible} SM particles, i.e. not neutrinos.  Due to the tight constraints on proton decay, we do not consider operators which violate baryon number.  This limits our current work to only the processes $f d \rightarrow u e^-$ and $\bar{f} u \rightarrow d e^+$ and, thus, we concentrate on the process $\bar{f}p\rightarrow n e^+$.  Although one could consider processes with both visible SM particles and additional invisible, non-SM particles in the final state, we do not consider that case here.  Thus, we consider operators which contain exactly one non-SM $f$ field.  We eliminate redundant operators using integration by parts and the equations of motion for the fields; although $f$ may have mass or Yukawa-like couplings to other non-SM particles, we can still use its equation of motion to eliminate operators containing $\slashed{D}f$ in favor of the operators appearing below or operators which do not contribute to the processes we are interested in here.

At dimension four, there is only one operator, $\bar{L}\tilde{\phi}f$; this operator contributes at tree level to neutrino mass, however, and is thus severely constrained.  All operators of dimension five can be eliminated by integration by parts and the equations of motion of the fields.  Thus, we consider operators of dimension six.  Again, we ignore operators which are constrained at tree level by neutrino mass.  With these restrictions, we are left with six operators
\begin{eqnarray}
{\cal O}_{W} &=& g \bar{L} \tau^a \tilde{\phi} \sigma^{\mu\nu} f W^a_{\mu\nu}\nonumber\\
{\cal O}_{\tilde{V}} &=& \bar{\ell}_R\gamma_{\mu}f \phi^{\dagger} D_{\mu} \tilde{\phi}\nonumber\\
{\cal O}_{VR} &=& \bar{\ell}_R \gamma_{\mu} f \bar{u}_R \gamma^{\mu} d_R, \\
{\cal O}_{Sd} &=& \epsilon_{ij }\bar{L}^i  f \bar{Q}^j d_R\nonumber\\
{\cal O}_{Su} &=& \bar{L} f \bar{u}_R Q \nonumber\\
{\cal O}_{T} &=& \epsilon_{ij }\bar{L}^i \sigma^{\mu\nu} f \bar{Q}^j \sigma_{\mu\nu} d_R .\nonumber
\end{eqnarray}
These operators are all suppressed by $\Lambda^2$, where $\Lambda$ is some mass taken to be higher than the weak scale, and each operator ${\cal O}_I$ is multiplied by a coefficient $C_I$.  Note that in all of these operators, $f$ is right-handed.  Although $f$ has the same quantum numbers as those of a right-handed neutrino, we do not, at this stage, require $f$ to have any specific interactions, such as a coupling to a right-handed $Z'$, other than those specified above.  For simplicity, we will neglect the possibility that $f$ interacts through more than one of the above operators simultaneously.

\section{Dark Matter Constraints}
\label{sec:dm}
Having enumerated the operators ${\cal O}_I$ which can contribute to $\bar{f} p \rightarrow e^+ n$, we now wish to place constraints on their coefficients divided by square of the new physics scale, $C_I/\Lambda^2$.  In order for $f$ to be DM, it must be sufficiently long lived; its lifetime must be at least of the order of the age of the universe ($\sim 4\times 10^{17}$ s), and it must decay to visible products sufficiently slowly to have not yet been detected.  We consider the constraints from DM lifetime on each of these operators in turn.
\begin{subsection}{${\cal O}_{W}$}
${\cal O}_{W}$ contributes to the decay $f\rightarrow \nu \gamma$ at tree level;
\begin{equation}
\Gamma(f\rightarrow\nu\gamma) = \frac{|C_W|^2}{\Lambda^4} \frac{\alpha v^2}{2} m_f^3
\end{equation}
where $v\sim 246$ GeV is the Higgs vacuum expectation value.  In order for $f$ to have a lifetime comparable to the age of the universe, $\Gamma \lesssim (4 \times 10^{17} \mbox{s})^{-1}=1.6\times 10^{-42}$ GeV.  However, for $f$ to be observable via the processes  $f n \rightarrow p e^-$ or $\bar{f} p \rightarrow e^+ n$ at neutrino experiments, its mass must be at least a significant fraction of an MeV \cite{Amsler:2008zzb}.  We obtain 
\begin{equation}
\label{eq:ownum1}
\frac{|C_W|^2}{\Lambda^4}\lesssim \frac{1}{(6\times 10^5 \mbox{TeV})^4}\left(\frac{1\mbox{ MeV}}{m_f}\right)^3.
\end{equation} 

The insistence of a long lifetime for DM is clearly quite constraining.  However, an even stronger limit on $\frac{|C_W|^2}{\Lambda^4}$ is possible, using the results of \cite{Yuksel:2007dr}, who use limits on $\gamma$-ray emission from the galactic center determined by INTEGRAL \cite{Teegarden:2006ni} and measurements of the diffuse $\gamma$-ray background from INTEGRAL \cite{Churazov:2006bk}, COMPTEL \cite{Weidenspointner:2000ab} and EGRET \cite{Sreekumar:1997un,Strong:2004ry} to constrain the lifetime of DM decaying to two daughter particles, one of which is a photon.  They find that the DM lifetime must be $\gtrsim O(10^{26}\mbox{ s})$ for a DM mass between $\sim 1$ MeV and $\sim 100$ GeV, with even tighter constraints toward the lower end of this range and down to $.04$ MeV.  Thus, we obtain a limit on $\frac{|C_W|^2}{\Lambda^4}$ more than two orders of magnitude stronger than that in Eq. (\ref{eq:ownum1}), $\sim 1/(8 \times 10^7 \mbox{ TeV})^4 (1\mbox{ MeV}/m_f)^3$, a scale clearly inaccessible at neutrino experiments or collider experiments such as LHC. 

We note in passing that, although not interesting for current neutrino experiments, masses as low as $0.4$ keV are allowed for fermionic dark matter \cite{Boyarsky:2008ju}.  However, limits from \cite{Kawasaki:1997ah} constrain the lifetime of DM particles to be orders of magnitude longer than the age of the universe for DM masses as low as $0.2$ keV,  implying that for even very light $O(\mbox{keV})$ DM, the effects of ${\cal O}_W$ would be undetectable at both neutrino experiments and at LHC. 
\end{subsection}

\begin{subsection}{${\cal O}_{\tilde{V}}$}
After Electroweak Symmetry Breaking the Higgs field $\phi$ obtains a vacuum expectation value, and ${\cal O}_{\tilde{V}}$ takes on the effective form 
\begin{equation}
\frac{C_{\tilde{V}}}{\Lambda^2}{\cal O}_{\tilde{V}} \rightarrow \frac{-ig C_{\tilde{V}}v^2}{2\sqrt{2}\Lambda^2} \bar{\ell}_R \gamma^{\mu} f W_{\mu}^-.
\end{equation}
If $m_f\gtrsim 2 m_e\sim 1$ MeV, this allows the decay $f\rightarrow e^+ e^- \nu$ at tree level.  For the range of $m_f$ valid for the constraints from Super-Kamiokande which we will consider in section \ref{sec:nus}, the electron and positron masses can be neglected, and ${\cal O}_{\tilde{V}}$  gives the $f$ a decay width of
\begin{equation}
\Gamma(f\rightarrow e^+ e^- \nu)=\frac{|C_{\tilde{V}}|^2}{\Lambda^4} \frac{1}{1536 \pi^3} m_f^5. 
\end{equation}
Requiring this to be less than $1.6\times 10^{-42}$ GeV and taking $m_f\sim$ few MeV, we obtain
\begin{equation}
\frac{|C_{\tilde{V}}|^2}{\Lambda^4} \lesssim O\left(\frac{1}{(10^3\mbox{ TeV})^4}\right),
\end{equation}
with even stronger limits for larger values of $m_f$.  Thus, we see that the effects of ${\cal O}_{\tilde{V}}$ are out of reach for current-day neutrino and collider experiments for $m_f\gtrsim 1$ MeV.  However, even stronger limits are possible; \cite{Picciotto:2004rp} compute the decay width of an arbitrary DM particle to $e^+e^-$ using the 511 keV line measured by INTEGRAL.  For the case where $f$ comprises all DM, and where its decays are responsible for the 511 keV line, their result corresponds to a lifetime $\tau_{\tilde{V}}$ of
\begin{equation}
\tau_{\tilde{V}} \simeq 5\times 10^{17} \mbox{yr} \frac{10 \mbox{ MeV}}{m_f}.
\end{equation}
This  gives\footnote{For a slightly weaker bound using decays involving neutrinos, see \cite{PalomaresRuiz:2007ry}.}
\begin{eqnarray}
\frac{|C_{\tilde{V}}|^2}{\Lambda^4} &\lesssim& \frac{1}{(9.5\times 10^5\mbox{ TeV})^4} \mbox{ ($m_f=20$ MeV)}\nonumber\\
&\lesssim&\frac{1}{(2.4\times 10^6 \mbox{ TeV})^4} \mbox{ ($m_f=50$ MeV)}\\
&\lesssim&\frac{1}{(3.8\times 10^6\mbox{ TeV})^4} \mbox{ ($m_f=80$ MeV)},\nonumber
\end{eqnarray}
from which one can see a linear dependence of the NP scale on $m_f$.  We note, however, that, taken as limits on the new physics scale, these numbers are probably rather conservative.  As discussed briefly below, \cite{Beacom:2005qv} and \cite{Sizun:2006uh} conclude that the 511 keV line can be produced only by processes in which the injection energy of the positrons was less than $\sim 3$ or $\sim 7.5$ MeV, respectively.  Therefore, we note that, for example, if $f$ decays via ${\cal O}_{\tilde{V}}$ were responsible for $10\%$ of the 511 keV flux, this would tighten the limits on the scale of new physics $\Lambda$ by a factor of $\sim 1.8$.  

We will now use the very tight constraints for ${\cal O}_{\tilde{V}}$ and ${\cal O}_W$ to place constraints on the remaining operators.  
\end{subsection}
\begin{subsection}{${\cal O}_{VR}$}
If $m_f>m_{\pi} + m_e$, ${\cal O}_{VR}$ gives the decay $f\rightarrow \pi^+ e^-$ at tree level.  Thus, we consider the case where $m_f\lesssim m_{\pi}$; here, $f$ can decay as $f\rightarrow e^+e^-\nu$ via the mixing of ${\cal O}_{VR}$ into ${\cal O}_{\tilde{V}}$ shown in the diagram in Fig.~\ref{fig:oneloop}.  We note that both quark fields in ${\cal O}_{VR}$ are right-handed.  Thus, the mixing of ${\cal O}_{VR}$ into ${\cal O}_{\tilde{V}}$ must be suppressed by the $u$ and $d$ masses as both quarks must flip chirality to interact with the $W$.  The diagram in Fig. \ref{fig:oneloop} is logarithmically divergent; we thus estimate the mixing of ${\cal O}_{VR}$ into ${\cal O}_{\tilde{V}}$ as
\begin{equation}
\frac{C_{\tilde{V}}(v)}{\Lambda^2}\sim\frac{C_{VR}(\Lambda)}{\Lambda^2}\frac{1}{(4\pi)^2}\frac{12 m_u m_d}{v^2}\ln{\frac{\Lambda^2}{m_f^2}}.
\label{eq:ovr}
\end{equation}
We obtain results of 
\begin{eqnarray}
\frac{|C_{VR}|^2}{\Lambda^4} &\lesssim& \frac{1}{(20\mbox{ TeV})^4} \mbox{ ($m_f=20$ MeV)}\nonumber\\
&\lesssim&\frac{1}{(50 \mbox{ TeV})^4} \mbox{ ($m_f=50$ MeV)}\label{eq:ovr15}
\\
&\lesssim&\frac{1}{(80\mbox{ TeV})^4} \mbox{ ($m_f=80$ MeV)}\nonumber
\end{eqnarray}
for $m_f$ in the range which will be relevant for the limits from Super-K discussed in section \ref{sec:nus}.  Here, like in the case of ${\cal O}_{\tilde{V}}$, we see a linear dependence of the NP scale on $m_f$.  We have taken the values of the light quark masses from \cite{Amsler:2008zzb}; given the uncertainty in these quantities, the error on our estimates of the NP scale should be taken as on the order of $30\%$.
\begin{figure}[h]
\includegraphics[width=.5\textwidth, angle=0]{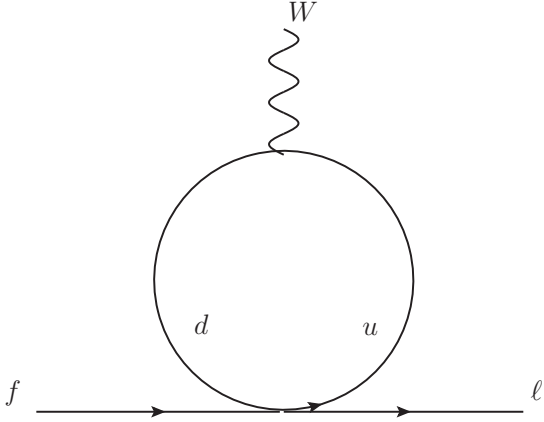}
\caption{Mixing of ${\cal O}_{VR}$ into ${\cal O}_{\tilde{V}}$.}
\label{fig:oneloop}
\end{figure}

However, this expression does not accurately describe contributions from small ($\lesssim \mbox{ few }\times 100 \mbox{ MeV}$) internal loop momenta.  For this region, we instead calculate the contribution from a diagram where the internal $u$ and $d$ quarks are replaced by a virtual $\pi^+$.   The amplitude squared for this diagram is
\begin{equation}
\label{eq:ovr2}
\frac{1}{2}\displaystyle\sum_{spins}|{\cal M}|^2= \frac{|C_{VR}|^2}{\Lambda^4} G_F^2 |V_{ud}|^2 f_{\pi}^4 m_e^2 m_f^2\left(\frac{1}{4}\right) \frac{q^2(m_f^2-q^2)}{(q^2-m_{\pi}^2)^2}
\end{equation}
where $G_F$ is the Fermi constant, $q=p_f-p_e$, and $f_{\pi}$ is the pion decay constant, $\sim 130$ MeV.  Note that this expression contains factors of $m_f$ and $m_e$, analogous to the lepton mass dependence that results in the usual SM $\pi^+$ decay helicity suppression.  Thus, if the decay channel $f\rightarrow e^- \mu^+ \nu_{\mu}$ is open, the constraints on $C_{VR}/\Lambda^2$ will be substantially stronger.  However, for $m_f\lesssim m_{\mu}$, the contribution to $f\rightarrow e^-e^+\nu$ from the diagram containing a virtual $\pi^+$ gives the differential cross-section
\begin{equation}
\frac{d\Gamma}{dq^2}=\frac{|C_{VR}|^2}{\Lambda^4}\frac{G_F^2|V_{ud}|^2f_{\pi}^4 m_e^2}{1024 \pi^3 m_f }\frac{q^2(m_f^2-q^2)^2}{(q^2-m_{\pi}^2)^2}
\end{equation}
which yields the limits
\begin{eqnarray}
\frac{|C_{VR}|^2}{\Lambda^4} &\lesssim& \frac{1}{(6\mbox{ TeV})^4} \mbox{ ($m_f=20$ MeV)}\nonumber\\
&\lesssim&\frac{1}{(20 \mbox{ TeV})^4} \mbox{ ($m_f=50$ MeV)}\\
&\lesssim&\frac{1}{(50\mbox{ TeV})^4} \mbox{ ($m_f=80$ MeV)}.\nonumber
\label{eq:ovr2.5}
\end{eqnarray}
These results are somewhat less restrictive than the result obtained via the loop diagram in Eq. (\ref{eq:ovr15}).  Thus, we take our results in Eq. (\ref{eq:ovr15}) as our approximte limits on the NP scale.

Given that ${\cal O}_{VR}$ is substantially more weakly constrained than the previous two operators, it is fruitful to look for other possible constraints.  First, we consider the possible decay $f\rightarrow \gamma\nu$ via the mixing of ${\cal O}_{VR}$ into ${\cal O}_{W}$.  We note that ${\cal O}_{VR}$ does not mix into ${\cal O}_{W}$ at one-loop order.  Example diagrams of the contribution to  $f\rightarrow \nu \gamma$ at two-loop order are shown in Fig. \ref{fig:twoloop}; however, this contribution is much weaker than the one-loop result for $f\rightarrow e^+e^-\nu$, due to an extra suppression from the electron mass; in Fig. \ref{fig:twoloop} (a), this is due to an electron chirality flip, while in Fig. \ref{fig:twoloop} (b), it is due to a coupling of the electron to the Higgs field.

We also estimate a limit on the NP scale by the contribution of ${\cal O}_{VR}$ to $f$ decay via $f-\nu$ mixing.  This can occur through the diagrams similar to those in Fig. \ref{fig:twoloop}, but with the external photon removed.  The diagram similar to that in Fig. \ref{fig:twoloop} (b) is quadratically divergent, and, thus, the contribution to an effective $\bar{\nu}_L f$ mass term is of order
\begin{equation}
\frac{C_{VR}}{(4\pi)^4} \frac{2 m_e m_u m_d }{v^2} \sim 10^{-8} \mbox{ eV}
\label{eq:numix}
\end{equation}
as we have cut off the quadratically divergent integral at $O(\Lambda)$ and taken $C_{VR}\sim 1$.  We expect the mixing angle between the $f$ and $\nu$ to be of order $10^{-8}\mbox{ eV}/m_f\sim 10^{-16}$.  This mixing allows the decays $f\rightarrow \nu \nu\bar{\nu}$ and $f\rightarrow \nu e^+ e^-$ via couplings to the SM $Z$ or $W$, suppressed by the square of the mixing angle.  Thus, we expect the lifetime of the $f$ to be of order $\tau_{\mu}\times 10^{32}\sim 10^{26}\mbox{ s}$ (where $\tau_{\mu}$ is the $\mu$ lifetime), longer than the requirement stated earlier of $\sim 10^{17}\mbox{ yr}$.

Finally, for comparison, we can also consider constraints from $\pi$ decay, as ${\cal O}_{VR}$ would contribute to $\pi^+ \rightarrow e^+ f$.  Searches for heavy neutrinos in $\pi$ decay for $m_f\lesssim 130$ MeV \cite{Britton:1992xv} give constraints on $C_{VR}/\Lambda^2$ of order $1/(10 \mbox{ TeV})^2$.
\begin{figure}[h]
\includegraphics[width=1\textwidth, angle=0]{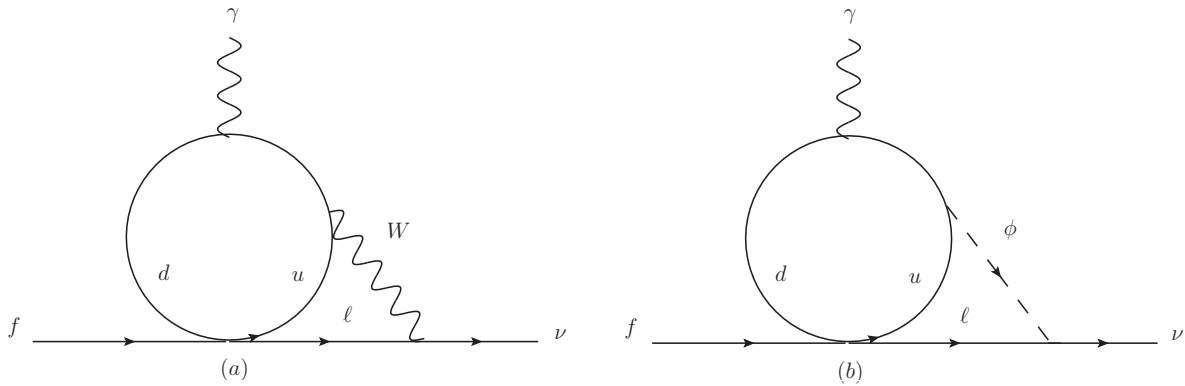}
\caption{Example diagrams for mixing of ${\cal O}_{VR}$, ${\cal O}_{Sd}$, ${\cal O}_{Su}$ into ${\cal O}_{W}$ at two-loop order.}
\label{fig:twoloop}
\end{figure}

\end{subsection}
\begin{subsection}{${\cal O}_{Sd}$, ${\cal O}_{Su}$, and ${\cal O}_{T}$}
Operators ${\cal O}_{Sd}$ and ${\cal O}_{Su}$ do not mix into ${\cal O}_{W}$ or ${\cal O}_{\tilde{V}}$ at one-loop order.  They can be inserted into diagrams like that in Fig. \ref{fig:twoloop}(a) to give $f\rightarrow \nu \gamma$.  (Diagrams which contain a Higgs, such as in Fig. \ref{fig:twoloop}(b), will be suppressed by additional powers of small Yukawa couplings.)  As these operators contain only one right-handed SM field, there are contributions with suppression by only one power of a light mass; we give an order-of-magnitude estimate for the mixing into ${\cal O}_W$:
\begin{equation}
\frac{C_W(v)}{\Lambda^2}\sim \frac{C_{Su,Sd}(\Lambda)}{\Lambda^2} \frac{1}{(4 \pi)^4} g^2 \frac{m_{u,d}}{v} \ln{\left(\frac{\Lambda^2}{v^2}\right)}\sim \frac{C_{Su,Sd}(\Lambda)}{\Lambda^2}\times 10^{-9}.
\label{eq:osu}
\end{equation}
For $m_f=1$ MeV, in order for $f$ to satisfy the $\gamma$-ray constraints, we must have
\begin{equation}
\frac{C_W(v)}{\Lambda^2}\lesssim \frac{1}{(8\times 10^7\mbox{ TeV})^2}
\end{equation} 
with tighter constraints for larger values of $m_f$.  This implies the order-of-magnitude limit
\begin{equation}
\frac{C_{Su,Sd}(\Lambda)}{\Lambda^2} < O\left(\frac{1}{(10^3 \mbox{ TeV})^2}\right).
\end{equation}

${\cal O}_T$, on the other hand, mixes into ${\cal O}_W$ at one-loop order and will be more strongly constrained than ${\cal O}_{Su}$ and ${\cal O}_{Sd}$.  We thus find that ${\cal O}_{Su}$, ${\cal O}_{Sd}$, and ${\cal O}_T$ to be very tightly constrained, likely beyond the reach of near-future experiments.

\end{subsection}

Given the above constraints, we concentrate for the rest of this work on the operator ${\cal O}_{VR}$ and restrict ourselves to the mass range $m_f\lesssim m_{\pi}$.  It is appropriate at this stage to review the motivation for, and the existing constraints on, DM with a mass of order $1-100$ MeV.  Although some supersymmetry-inspired models of MeV DM exist \cite{Hooper:2008im,Boehm:2003ha}, its original motivation has been the proposal \cite{Boehm:2002yz,Boehm:2003hm,Boehm:2003bt} that MeV DM annihilating to $e^+e^-$ could explain the 511-keV line from the galactic bulge observed by INTEGRAL \cite{Knodlseder:2005yq,Jean:2005af}.  It has also been argued \cite{Hooper:2003sh} that searches for 511-keV emissions from DM-rich dwarf spheriodals \cite{Cordier:2004hf} would provide very compelling evidence for annihilating DM.  However, it has been shown \cite{Beacom:2005qv} that in-flight positron annihilations on electrons in the intersteller medium would over-produce galactic $\gamma$ rays unless their injection energy was $\lesssim 3$ MeV, severely restricting the range of DM masses compatible with the 511 keV line. A similar analysis \cite{Sizun:2006uh} obtained an upper limit of $7.5$ MeV; see also \cite{Beacom:2004pe,Ascasibar:2005rw,Boehm:2006df}. However, while this makes the case for MeV-scale DM less observationally motivated, it merely constrains the annihilation of DM to $e^+e^-$ at late times, not the {\it existence} of MeV-scale DM.  Therefore, we will relax the requirement that MeV-scale DM be responsible for the observed 511-keV line, and consider the entire possible mass range $m_f\lesssim m_{\pi}$.

With any DM candidate, its present-day relic density must be compatible with observation.  The interaction ${\cal O}_{VR}$ is similar to that for a right-handed neutrino.  As is well-known, the Lee-Weinberg bound \cite{Lee:1977ua} requires that heavy neutrinos which annihilate through weak-scale intermediate gauge bosons have a mass $\gtrsim O(\mbox{GeV})$ in order to annhilate sufficiently to not overclose the universe.  Interactions suppressed by higher physics scales will freezeout earlier; if these interactions dominate $f$ annihilations, $f$ will annihilate less completely, thus strengthening the Lee-Weinberg bound.  Thus, in order for $f$ to be a realistic DM candidate with $m_f\lesssim m_{\pi}$, it must have at least one other interaction in addition to ${\cal O}_{VR}$ which is stronger than weak interactions.

A number of works \cite{Boehm:2002yz,Boehm:2003bt,Fayet:2004bw,Rasera:2005sa,Serpico:2004nm,Boehm:2003hm,Schleicher:2008gm,Boehm:2004gt} have studied interactions which can make the relic density of MeV-scale DM agree with observation.  If we assume that $f$ dominates the dark matter density, we must have a thermally-averaged cross-section for annihilation $\langle \sigma_{ann}|v_{r}|\rangle$ at freezeout of $O(10^{-25}) \mbox{cm}^3/\mbox{s}$ \cite{Boehm:2002yz,Boehm:2003bt,Boehm:2003ha,Fayet:2004bw,Serpico:2004nm,Ahn:2005ck}, where $v_r$ is the relative velocity of DM particles.  However, if the primary annihilation channel during freezeout is $f\bar{f}\rightarrow e^+e^-$, there is tension between the cross-section implied by the present-day relic density and the flux for the 511 keV line, if the cross-section is independent of velocity, i.e., s-wave \cite{Boehm:2003bt,Boehm:2003ha,Hooper:2003sh,Fayet:2004bw}; similar conclusions hold for gamma rays over a range of energies \cite{Serpico:2004nm,Schleicher:2008gm,Bell:2008vx}.  However, if the cross-section is p-wave, with $\langle \sigma_{ann}|v_{r}|\rangle\sim v_r^2$, these requirements can be brought into better agreement; \cite{Boehm:2003hm,Boehm:2003bt,Boehm:2004gt,Fayet:2004bw,Fayet:2007ua, Fayet:2006xd} have suggested obtaining such behavior by coupling both DM particles and $e^+e^-$ to a new, light U-boson.  \cite{Fayet:2004bw} find that such a U-boson with an axial-vector coupling to a DM fermion, such as occurs for Majorana DM particles, will either give a cross-section that is completely p-wave (for a vector coupling to electrons) or whose s-wave component is suppressed by the electron mass (for an axial vector coupling to the electron).\footnote{There have been many studies of the phenomenology of a U-boson in high-energy colliders, low-energy colliders \cite{McElrath:2005bp,Borodatchenkova:2005ct,Ablikim:2006eg,Zhu:2007zt,Fayet:2008cn,Fayet:2006sp}, neutrino physics \cite{Boehm:2004uq}, and its contribution to the electron magnetic moment \cite{Boehm:2007na,Boehm:2004gt}.  For reviews of U-boson detectability, see \cite{Fayet:2007ua,Fayet:2006xd}.}  

Here, as we wish to be model-independent, we will stick with the effective operator formalism; for example, for a velocity-dependent annhilation to $e^+e^-$, we could introduce the operator
\begin{equation}
\frac{C_{Ve}}{\Lambda_a^2} {\cal O}_{Ve}= \frac{C_{Ve}}{\Lambda_a^2}\bar{f}\gamma^{\mu}\gamma_5 f \bar{\ell}_{R}\gamma_{\mu}\ell_{R}
\end{equation}
where the $\ell_R$ fields are taken to be $e$ flavor.  Here, we do not take $\Lambda_a$ to be similar to the NP scale $\Lambda$ for ${\cal O}_{VR}$; we take ${\cal O}_{Ve}$ to be an effective operator valid at energies relevant for freezeout, $O(m_f)$, although one could consider processes mediated by intermediate particles which are sufficiently light that this approximation is not valid, if the intermediate particle, such as a U-boson, is very weakly coupled.  We leave the investigation of how such an operator may arise for future work; we make a few brief comments on model-building in section \ref{sec:collflav}.  Taking $v_r\sim 0.4 c$ at freezeout \cite{Fayet:2004bw}, and requiring the thermally-averaged annihilation cross-section $\langle \sigma_{ann}|v_{r}|\rangle$ to be $O(10^{-25} \mbox{ cm}^3/\mbox{s})$, $|C_{Ve}|/\Lambda_a^2$ is of order $1/(\mbox{few GeV})^2$ for $f$ in the mass range of interest here\footnote{We note that if the physics which mediates SM-DM interactions contains very weak couplings, this GeV-scale operator could correspond to intermediate particles lighter than a GeV.}.

However, given the results of \cite{Beacom:2005qv} and \cite{Sizun:2006uh} which show that the 511 keV line can only be produced by annihilations of DM with very low mass, considering only annihlations to $e^+e^-$ is not well-motivated.  Thus, we also consider the case where $f$ annihilates primarily to neutrinos.  We can do this by introducing the operator 
\begin{equation}
\frac{C_{V\tau}}{\Lambda_a^2} {\cal O}_{V\tau}= \frac{C_{V\tau}}{\Lambda_a^2}\bar{f}_R\gamma^{\mu}f_R \bar{L}_{\tau}\gamma_{\mu}L_{\tau}
\end{equation} 
or the same operator where the lepton fields are $\mu$ flavor.  This operator allows the annihilations $f\bar{f}$ to both $\tau^+\tau^-$ and $\nu_{\tau}\bar{\nu_{\tau}}$.  However, for the mass range of interest here, $m_f\lesssim 140$ MeV, only the neutrino channel is open both during freezeout and for present-day annihilations; in the case of the $\mu$-flavored operator, for $m_f>m_{\mu}$, annihilations to muons may form some fraction of the cross-section \cite{Bell:2008vx}.  As before, relic density considerations require $\langle \sigma_{ann}|v_{r}|\rangle$ to be of order $10^{-25} \mbox{ cm}^3/\mbox{s}$.  Such an operator could again arise from the exchange of a light U-boson; however, keeping to our model-independent philosophy, here we again only assume it to be some effective operator valid for energies $O(m_f)$.  As ${\cal O}_{V\tau}$ does not contribute to the 511 keV line, a velocity-independent cross-section is acceptable; optionally, we could then add in other operators which could contribute to, but not fully explain, the 511 keV line, as long as they satisfy the requirements imposed by \cite{Beacom:2005qv,Sizun:2006uh,Schleicher:2008gm}.  Upper limits on the velocity-averaged annihilation cross-section of DM to neutrinos for DM mass above \cite{Yuksel:2007ac} and below \cite{PalomaresRuiz:2007eu} $\sim100$ MeV are $\mbox{few}\times(10^{-24}\mbox{ cm}^3/\mbox{s})$ and $O(10^{-25}\mbox{ cm}^3/\mbox{s})$, respectively.  While these limits are not far from the required value for $\langle \sigma_{ann}|v_{r}|\rangle$, this annihilation channel is not definitively ruled out.  Again, taking $<\sigma_{ann} |v_r|>\sim O(10^{-25}\mbox{ cm}^3/\mbox{s})$ gives a physics scale for this operator of order several GeV.   
 
There are also constraints on MeV-scale DM from supernovae.  \cite{Barbieri:1988av} placed limits on the interactions of right-handed neutrinos by requiring that they interact sufficiently strongly to become trapped and not cool the supernova too rapidly, contradicting observations from SN 1987A.  Here, in order to satisfy the relic density requirements, we assume that $f$ interacts with electrons and/or neutrinos with a strength much stronger than the weak interactions; in this case, the $f$ is trapped.  However, if the interaction of the $f$ with neutrinos is too strong, it can make the neutrinos more trapped and cause the supernova to cool too slowly; \cite{Fayet:2006sa} consider the case where a DM particle is strongly coupled to neutrinos and conclude that such a DM particle is ruled out if it has a mass $\lesssim 10$ MeV and that this lower bound could be raised to as much as $\sim30$ MeV.

The effects of MeV DM on big-bang nucleosynthesis (BBN) have also been studied.  \cite{Serpico:2004nm} consider DM coupled to neutrinos, electrons, and to both neutrinos and electrons simultaneously.   They find, for the neutrino case, that DM particles of mass $\gtrsim 10$ MeV pose no problem for BBN.  For the case where DM is coupled to electrons, they find that the cross-section necessary for the correct relic density is incompatible with BBN if the interaction is s-wave and the DM mass is between $20$ and $90$ MeV.  P-wave interactions, however, were still compatible.  For the case where the DM particle was coupled to both electrons and neutrinos, they expected to have no discrepancy with BBN for a DM mass $\gtrsim 20$ MeV. 

Finally, we note that the effects of MeV DM for $m_f\lesssim 3$ MeV on the formation of small-scale structure has been studied in \cite{Hooper:2007tu}.

Thus, we see that fermionic DM with a mass $O(10-100\mbox{ MeV})$ is not definitively ruled out by any of the above considerations, although further investigation would require the formulation of a specific model.  We now move on to calculating the interaction rate of such DM particles in Super-K.

\section{Signals in Super-Kamiokande}
\label{sec:nus}
We now discuss how the $f$ particle could be observed in neutrino experiments.  For this purpose, we will focus on the relic supernova neutrino search using the process $\bar{\nu}_e p \rightarrow n e^+$ from Super-Kamiokande \cite{Malek:2002ns}.  We consider this analysis as it covers a wide range of neutrino energy $E_{\nu}$, and this range in $E_{\nu}$ can be translated into a range of possible values of $m_f$ in the process $\bar{f} p \rightarrow n e^+$.  Super-Kamiokande has placed a limit of $1.2$ $\bar{\nu}_e / \mbox {cm}^2 \mbox{s}$ on the relic antineutrino flux for $19.3 \mbox{ MeV} < E_{\nu} \lesssim 80 \mbox{ MeV}$.  They fit the overall normalization of their backgrounds to data, as the energy distributions for their backgrounds and for their signal are sufficiently distinct; as the electron in the process $\bar{f} p \rightarrow n e^+$ should be approximately monoenergetic with an energy of $m_f$ (before being modulated by the detector resolution), we assume that this is a conservative estimate of how well this process could be distinguished from background.  Although it is possible that the difference between the shapes of the $\bar{\nu}_e$ and $\bar{f}$ energy distributions could result in our obtaining an overly optimistic limit on $\bar{f} p \rightarrow n e^+$ for specific values of $m_f$, we would expect this would change the limit on the scale of new physics by at most a few tens of percent. 

To calculate the interaction rate of $\bar{f}$ at Super-K, the first thing which we must know is the DM flux.  The mass density of DM in our neighborhood is $\sim 0.3 \mbox{ GeV}/\mbox{cm}^3$ \cite{Caldwell:1981rj}, and its velocity $v_f$ relative to the Earth is $O(10^{-3})$ (for $c=1$)\cite{Kamionkowski:1997xg}.  Thus, assuming that $\bar{f}$ constitutes all DM, the $\bar{f}$ flux $\Phi_{\bar{f}}$ is
\begin{equation}
\Phi_{\bar{f}}\sim\frac{0.3\mbox{ GeV}/\mbox{cm}^3}{m_f} v_f c \sim (10^{10},10^9,10^8) /\mbox{cm}^2\mbox{s}
\end{equation}
for $m_f=(1, 10, 100)$ MeV, respectively.  Thus, Super-K can be sensitive to the process $\bar{f} p \rightarrow n e^+$ as long as its cross-section $\sigma_{{\cal O}}$ is $\gtrsim 10^{-8}$ times the SM neutrino cross-section evaluated at the neutrino energy $E_{\nu} = m_f$.  The relevant $\bar{f}$ flux and corresponding sensitivity to  $\sigma_{{\cal O}}$ are reduced by a factor of $2$ if $f$ and $\bar{f}$ are present in equal numbers. 

Next, we calculate the ratio of $\sigma_{{\cal O}}$ to the SM $\bar{\nu}_e$ cross-section.  For simplicity, we neglect the electron mass, the proton-neutron mass difference, and the possibility of second-class currents.  We only include terms leading order in $m_f/M\lesssim 0.1$ (and $E_{\nu}/M$), where $M$ is the nucleon mass.  These approximations should not change our values for the NP scale by more than $O(10\%)$, which is adequate for our purposes.  The cross-section $\sigma_{{\cal O}}$ is 
\begin{equation}
\sigma_{{\cal O}}\simeq \frac{1}{16\pi |v_f|} \frac{ |C_{VR}|^2}{\Lambda^4}  m_f^2 (|f_1|^2+3|g_1|^2)
\end{equation}
where $f_1\simeq 1$ and $g_1\simeq -1.27$ are the nucelon form factors \cite{Strumia:2003zx}.  For the SM process $\bar{\nu}_e p \rightarrow n e^+$ we obtain, using expressions from \cite{Strumia:2003zx},
\begin{equation}
\sigma_{SM} \simeq \frac{G_F^2}{\pi} E_{\nu}^2 (|f_1|^2+3|g_1|^2).
\end{equation}
If we take the ratio of these two cross-sections with $E_{\nu}$ evaluated at $m_f$ and insist that this be less than $1.2/(\mbox{cm}^2s)/\Phi_{\bar{f}}$, we find 
\begin{equation}
\frac{|C_{VR}|^2 v^4}{8 |v_f| \Lambda^4}\leq \frac{1.2/\mbox{cm}^2\mbox{s}}{(0.3\mbox{ GeV}/\mbox{cm}^3) |v_f| c/m_f},
\end{equation}
where, again, $v$ is the Higgs vev and $c$ is in units of $\mbox{cm}/\mbox{s}$.  This gives 
\begin{eqnarray}
\label{eq:results}
\frac{|C_{VR}|^2}{\Lambda^4} &\lesssim& \frac{1}{(120\mbox{ TeV})^4} \mbox{ ($m_f=20$ MeV)}\nonumber\\
&\lesssim&\frac{1}{(90 \mbox{ TeV})^4} \mbox{ ($m_f=50$ MeV)}\\
&\lesssim&\frac{1}{(80\mbox{ TeV})^4} \mbox{ ($m_f=80$ MeV)}.\nonumber
\end{eqnarray}
Here, the accessible NP scale is proportional to $m_f^{-1/4}$ due to the dependence of the DM flux on $m_f$.  Thus, we see that the reach of neutrino experiments can be quite strong; for the mass range considered here, it gives limits on the NP scale of order $100$ TeV. 

A few notes about other possibly relevant experiments are in order.  First, \cite{Frere:2006hp} uses a calculation almost identical to ours to place limits on interactions of right-handed neutrino DM using Ga \cite{Abdurashitov:2000va,Hampel:1998xg} and Cl experiments \cite{Cleveland:1998nv}.  However, they considered the mass range $m_f\lesssim 5$ MeV, which is dominated by the solar neutrino background; for this mass range, the limits on the new physics scale are thus substantially weaker.  We also note that limits weaker than those above exist on the $\bar{\nu}_e$ flux for energies below $20$ MeV and as low as $1.8$ MeV from Super-K \cite{Gando:2002ub}, KamLAND \cite{Eguchi:2003gg}, and the Borexino Counting Test Facility \cite{Balata:2006db}; the limits on the new physics scale for low values of $m_f$ from these experiments would be weaker than those obtained above by a factor approximately ranging between a few and an order of magnitude.  Also, we note that there exist limits on the relic supernova $\nu_e$ (as opposed to the $\bar{\nu}_e$ considered above) flux from the Sudbury Neutrino Observatory \cite{Aharmim:2006wq} and an analysis of the Super-K data \cite{Lunardini:2008xd} which are about two orders of magnitude weaker than that of the $\bar{\nu}_e$ flux.  Constraints derived from the $\nu_e$ flux limits are relevant for the cases where 1) $f$ is a Dirac fermion and we consider $f$ (and not $\bar{f}$ as considered above) as our DM candidate, and 2) $f$ is Majorana, in which case the flux limits on both $\nu_e$ and $\bar{\nu}_e$ are constraining.   We do not consider constraints derived from the $\nu_e$ flux here; however, as the cross-section in neutrino detectors is proportional to $\Lambda^{-4}$, we would naively expect these constraints to translate into limits on the NP scale which are weaker than those in Eq. (\ref{eq:results}) by a factor of a few.  Finally, we note that facilities such as DUSEL \cite{Raby:2008pd} or Hanohano \cite{Learned:2008zj} may improve somewhat on these limits in the future.

%\section{}
%\label{sec:}

\section{Discussion}
\label{sec:collflav}
Having given our main results in section \ref{sec:nus}, we now discuss a few issues related to our analysis.

We will first consider how our results change if we consider the same operators, but with fields which are not just from the first generation.  First, we will consider the effects of changing the flavor on the quark fields in operators ${\cal O}_{VR}$, ${\cal O}_{Sd}$, ${\cal O}_{Su}$, and ${\cal O}_{T}$.  Since nucleons are made of up and down quarks, we must have operators which contain quarks from the first generation in order to have contributions to $\bar{f}p\rightarrow n \ell^+$.  The right-handed quark fields in these operators must therefore be of the first generation, while the left-handed quark doublet fields can be of the second and third generation, as their lower components contain a down quark, suppressed by a Cabibbo-Kobayashi-Maskawa (CKM) matrix term.  Thus, in the case of ${\cal O}_{VR}$, other flavor structures are not relevant for this work.  On the other hand, if we change the flavor of the left-handed quark doublet in ${\cal O}_{Su}$, the limit which we derived by an estimate of the mixing of this operator into ${\cal O}_{W}$ is no longer valid; the logarithmically divergent part of Fig. \ref{fig:twoloop}(a) is zero by the GIM mechanism.  For  ${\cal O}_{Sd}$, however, this diagram is merely CKM-suppressed; the same is true of the one-loop mixing of  ${\cal O}_{T}$ into  ${\cal O}_{W}$.  It is possible, however, to arrange linear combinations of flavor structures for which these constraints no longer apply, but it would be unexpected that the NP scale for the unconstrained flavor combinations is substantially lower than that of the flavor diagonal case where both quark fields are from the first generation.

Next, we consider the effects of changing lepton flavor.  Our constraints from $f$ lifetime for operators  ${\cal O}_{W}$, ${\cal O}_{Sd}$, ${\cal O}_{Su}$, and ${\cal O}_{T}$ are independent of the lepton mass, except that for $\mu$- and $\tau$-flavored operators, we are interested in higher values of $m_f$; therefore, changing the lepton flavor from $e$ to $\mu$ or $\tau$ will only result in stronger constraints.  In the case of ${\cal O}_{\tilde{V}}$, if the lepton is $\mu$ flavor, $f$ can decay at tree level via $f\rightarrow \mu^- e^+ \nu_e$ as long as $m_f\gtrsim m_{\mu} + m_e$.  However, as ${\cal O}_{\tilde{V}}$ is purely charged-current, it can only contribute to neutrino experiments where a muon is produced in the final state, which, even neglecting detector thresholds, requires $m_f\gtrsim m_{\mu}$.  A similar argument holds for the $\tau$ case.  Thus, this operator is not interesting for neutrino experiments, except perhaps for a very small range of $m_f$, which we ignore here.  

Next, we consider ${\cal O}_{VR}$; for simplicity, we will take the lepton field to be $\mu$ flavor.  First, we note that $f$ can decay at tree level via $f\rightarrow \mu^- \pi^+$ if $m_f>m_{\mu}+m_{\pi}\sim 245$ MeV.  Thus we are interested in the mass range $m_{\mu}\lesssim m_f \lesssim 245$ MeV.  Our expression for the one-loop mixing of ${\cal O}_{VR}$ into ${\cal O}_{\tilde{V}}$ in Eq. (\ref{eq:ovr}) is independent of lepton flavor, so we expect the $\mu$ case to be similarly strongly constrained, except for a small mass range $m_f\sim m_{\mu}$.  However, there may be some tension with neutrino mixing, as the mixing term in Eq. (\ref{eq:numix}) is dependent upon the lepton mass.  The observability of $f$ in neutrino experiments for the mass range $m_{\mu}\lesssim m_f \lesssim m_{\mu} + m_{\pi}$ we leave for future work.  We do not address the mass range $m_f<m_{\mu}$ for this case, as this is not relevant for neutrino experiments.

Some comments are also in order about the possible applicability of ${\cal O}_{VR}$ to specific models.  Here, we have taken $f$ to be a singlet under the SM.  If we assume a single Higgs doublet as in the SM, the quantum numbers of the fields in ${\cal O}_{VR}$ dictate that $f$ has the same quantum numbers as a right-handed neutrino, and that the interaction $\bar{L}\tilde{\phi}f$ is not disallowed by any symmetry.  This is problematic, as this operator would allow $f$ to mix with neutrinos and decay via $f\rightarrow \nu\nu\bar{\nu}$.  In order to prevent this, we must introduce a second Higgs doublet and impose a symmetry which disallows $\bar{L}\tilde{\phi}f$.  Once this is done, interactions must be introduced to give $f$ an appropriate relic density.   A complete investigation of these points, including whether or not any of the effective operators mentioned in section \ref{sec:dm} to produce a reasonable relic density (or additional operators not considered here, such as interactions coupling $f$ to quarks, or interactions involving very light intermediate particles where the effective operator formalism is not valid) are generated at an appropriately low scale, is beyond the scope of this work.  We observe, however, that ${\cal O}_{VR}$ appears similar to a right-handed neutrino interaction; this may imply that left-right symmetric models could be an elightening place to begin constructing a model which contains ${\cal O}_{VR}$ as a low-energy approximation.  

Finally, another point worth mentioning is that our above limits on the operator coefficients assume that $f$ is a DM particle.  We have not made any statements on the observability of these operators, for example at LHC, if $f$ belongs to a hidden sector, but is not a constituent of DM.  Thus, the investigation of these interactions, including various flavor structures, at LHC could still be a worthy endeavour.

\section{Conclusions}
\label{sec:conc}

In this work, we have considered the possibility that a light $O(1-100 \mbox{ MeV})$ fermionic DM particle $f$ could be observed by direct detection in current-day neutrino experiments if it interacts inelastically through the process $\bar{f} p \rightarrow n e^+$.  We have approached this problem from a model-independent standpoint and constrained operators contributing to this process using the DM lifetime and its decays to $\gamma \nu$ and to $e^+ e^- \nu$, finding lower bounds on the new physics scale $\Lambda$ for most of the operators to be $\gtrsim 1000 \mbox{ TeV}$.  Finding the constraints on one operator ${\cal O}_{VR}$ substantially weaker than those on the other operators, we then used the results from the Super-K relic supernova neutrino search to place limits on ${\cal O}_{VR}$.  We find that Super-K was able to probe the scale of new physics up to ${\cal O}(100 \mbox{ TeV})$, far beyond the reach of present-day or near-future colliders.

Given the impressive reach obtainable with Super-K, it is worthwhile to consider possible extensions of this work not explored here.  First, we have considered only charged-current interactions with nuclei; neutral-current interactions such as $f_1 p \rightarrow f_2 p$ and interactions with electrons could also be considered.  Also, if neutral-current interactions are considered, it is possible that, in addition to neutrino experiments, current direct-detection DM experiments designed to look for DM with masses of $O(\mbox{GeV-TeV})$ could be sensitive to much lighter DM interacting inelastically; the relevance of such interactions to DAMA/NaI has been explored in \cite{Bernabei:2008mv}.  Finally, we have only considered fermionic DM; one could also consider the cases of light dark scalars or light dark gauge bosons.

The main point of our paper, however, is that inelastic interactions in which a DM particle annihilates into a particle with a smaller mass (either an SM particle or another as-yet-unknown particle) can occur, and that these interactions can be very sensitively probed by existing experiments.  As we still know very little of the identity, or identities, of DM or how it interacts with itself or with ``regular'' matter, we should consider these possible interactions and investigate how existing constraints from neutrino experiments or experiments searching for elastically-interacting DM can be translated into constraints on more general scenarios.

\section{Acknowledgements}
The authors would like to thank M. Wise for his calculation of the decay width of $f\rightarrow e^+e^-\nu$, as well as numerous helpful comments and suggestions.  They would also like to thank H. Davoudiasl, S. Dawson, S. Gopalakrishna, W. Marciano, C. Sturm, and M. Ramsey-Musolf for extensive discussions and helpful advice.  This work is supported under US DOE contract No. DE-AC02-98CH10886.
\bibliographystyle{h-physrev}
%\bibliography{topbib}

\begin{thebibliography}{99}

%\cite{Bertone:2004pz}
\bibitem{Bertone:2004pz}
  G.~Bertone, D.~Hooper and J.~Silk,
  %``Particle dark matter: Evidence, candidates and constraints,''
  Phys.\ Rept.\  {\bf 405}, 279 (2005)
  [arXiv:hep-ph/0404175].
  %%CITATION = PRPLC,405,279;%%


%\cite{Strassler:2006im}
\bibitem{Strassler:2006im}
  M.~J.~Strassler and K.~M.~Zurek,
  %``Echoes of a hidden valley at hadron colliders,''
  Phys.\ Lett.\  B {\bf 651}, 374 (2007)
  [arXiv:hep-ph/0604261].
  %%CITATION = PHLTA,B651,374;%%

%\cite{Patt:2006fw}
\bibitem{Patt:2006fw}
  B.~Patt and F.~Wilczek,
  %``Higgs-field portal into hidden sectors,''
  arXiv:hep-ph/0605188.
  %%CITATION = HEP-PH/0605188;%%

%\cite{Han:2007ae}
\bibitem{Han:2007ae}
  T.~Han, Z.~Si, K.~M.~Zurek and M.~J.~Strassler,
  %``Phenomenology of Hidden Valleys at Hadron Colliders,''
  JHEP {\bf 0807}, 008 (2008)
  [arXiv:0712.2041 [hep-ph]].
  %%CITATION = JHEPA,0807,008;%%

%\cite{ArkaniHamed:2008qn}
\bibitem{ArkaniHamed:2008qn}
  N.~Arkani-Hamed, D.~P.~Finkbeiner, T.~R.~Slatyer and N.~Weiner,
  %``A Theory of Dark Matter,''
  Phys.\ Rev.\  D {\bf 79}, 015014 (2009)
  [arXiv:0810.0713 [hep-ph]].
  %%CITATION = PHRVA,D79,015014;%%

%\cite{Malek:2002ns}
\bibitem{Malek:2002ns}
  M.~Malek {\it et al.}  [Super-Kamiokande Collaboration],
  %``Search for supernova relic neutrinos at Super-Kamiokande,''
  Phys.\ Rev.\ Lett.\  {\bf 90}, 061101 (2003)
  [arXiv:hep-ex/0209028].
  %%CITATION = PRLTA,90,061101;%%




%\cite{TuckerSmith:2001hy}
\bibitem{TuckerSmith:2001hy}
  D.~Tucker-Smith and N.~Weiner,
  %``Inelastic dark matter,''
  Phys.\ Rev.\  D {\bf 64}, 043502 (2001)
  [arXiv:hep-ph/0101138].
  %%CITATION = PHRVA,D64,043502;%%

%\cite{Bernabei:2002qr}
\bibitem{Bernabei:2002qr}
  R.~Bernabei {\it et al.},
  %``Investigating The Dama Annual Modulation Data In The Framework Of Inelastic
  %Dark Matter,''
  Eur.\ Phys.\ J.\  C {\bf 23}, 61 (2002).
  %%CITATION = EPHJA,C23,61;%%

%\cite{TuckerSmith:2002af}
\bibitem{TuckerSmith:2002af}
  D.~Tucker-Smith and N.~Weiner,
  %``Inelastic dark matter at DAMA, CDMS and future experiments,''
  Nucl.\ Phys.\ Proc.\ Suppl.\  {\bf 124}, 197 (2003)
  [arXiv:astro-ph/0208403].
  %%CITATION = NUPHZ,124,197;%%

%\cite{TuckerSmith:2004jv}
\bibitem{TuckerSmith:2004jv}
  D.~Tucker-Smith and N.~Weiner,
  %``The status of inelastic dark matter,''
  Phys.\ Rev.\  D {\bf 72}, 063509 (2005)
  [arXiv:hep-ph/0402065].
  %%CITATION = PHRVA,D72,063509;%%


%\cite{Chang:2008gd}
\bibitem{Chang:2008gd}
  S.~Chang, G.~D.~Kribs, D.~Tucker-Smith and N.~Weiner,
  %``Inelastic Dark Matter in Light of DAMA/LIBRA,''
  arXiv:0807.2250 [hep-ph].
  %%CITATION = ARXIV:0807.2250;%%

%\cite{Cui:2009xq}
\bibitem{Cui:2009xq}
  Y.~Cui, D.~E.~Morrissey, D.~Poland and L.~Randall,
  %``Candidates for Inelastic Dark Matter,''
  arXiv:0901.0557 [hep-ph].
  %%CITATION = ARXIV:0901.0557;%%

%\cite{Alves:2009nf}
\bibitem{Alves:2009nf}
  D.~S.~M.~Alves, S.~R.~Behbahani, P.~Schuster and J.~G.~Wacker,
  %``Composite Inelastic Dark Matter,''
  arXiv:0903.3945 [hep-ph].
  %%CITATION = ARXIV:0903.3945;%%


%\cite{Finkbeiner:2009mi}
\bibitem{Finkbeiner:2009mi}
  D.~P.~Finkbeiner, T.~R.~Slatyer, N.~Weiner and I.~Yavin,
  %``PAMELA, DAMA, INTEGRAL and Signatures of Metastable Excited WIMPs,''
  arXiv:0903.1037 [hep-ph].
  %%CITATION = ARXIV:0903.1037;%%

%\cite{Batell:2009vb}
\bibitem{Batell:2009vb}
  B.~Batell, M.~Pospelov and A.~Ritz,
  %``Direct Detection of Multi-component Secluded WIMPs,''
  arXiv:0903.3396 [hep-ph].
  %%CITATION = ARXIV:0903.3396;%%

%\cite{Bernabei:2008mv}
\bibitem{Bernabei:2008mv}
  R.~Bernabei {\it et al.}  [DAMA Collaboration],
  %``Investigation on light dark matter,''
  Mod.\ Phys.\ Lett.\  A {\bf 23}, 2125 (2008)
  [arXiv:0802.4336 [astro-ph]].
  %%CITATION = MPLAE,A23,2125;%%




%\cite{Kim:2008pp}
\bibitem{Kim:2008pp}
  Y.~G.~Kim, K.~Y.~Lee and S.~Shin,
  %``Singlet fermionic dark matter,''
  JHEP {\bf 0805}, 100 (2008)
  [arXiv:0803.2932 [hep-ph]].
  %%CITATION = JHEPA,0805,100;%%

%\cite{Kim:2009ke}
\bibitem{Kim:2009ke}
  Y.~G.~Kim and S.~Shin,
  %``Singlet Fermionic Dark Matter explains DAMA signal,''
  arXiv:0901.2609 [hep-ph].
  %%CITATION = ARXIV:0901.2609;%%

%\cite{Lee:2008xy}
\bibitem{Lee:2008xy}
  K.~Y.~Lee, Y.~G.~Kim and S.~Shin,
  %``Singlet fermionic dark matter as a natural higgs portal model,''
  arXiv:0809.2745 [hep-ph].
  %%CITATION = ARXIV:0809.2745;%%

%\cite{Babu:2007sm}
\bibitem{Babu:2007sm}
  K.~S.~Babu and E.~Ma,
  %``Singlet fermion dark matter and electroweak baryogenesis with radiative
  %neutrino mass,''
  Int.\ J.\ Mod.\ Phys.\  A {\bf 23}, 1813 (2008)
  [arXiv:0708.3790 [hep-ph]].
  %%CITATION = IMPAE,A23,1813;%%

%\cite{Amsler:2008zzb}
\bibitem{Amsler:2008zzb}
  C.~Amsler {\it et al.}  [Particle Data Group],
  %``Review of particle physics,''
  Phys.\ Lett.\  B {\bf 667}, 1 (2008).
  %%CITATION = PHLTA,B667,1;%%



%\cite{Yuksel:2007dr}
\bibitem{Yuksel:2007dr}
  H.~Yuksel and M.~D.~Kistler,
  %``Dark Matter Might Decay... Just Not Today!,''
  Phys.\ Rev.\  D {\bf 78}, 023502 (2008)
  [arXiv:0711.2906 [astro-ph]].
  %%CITATION = PHRVA,D78,023502;%%

%\cite{Teegarden:2006ni}
\bibitem{Teegarden:2006ni}
  B.~J.~Teegarden and K.~Watanabe,
  %``A Comprehensive Search for Gamma-Ray Lines in the First Year of Data from
  %the INTEGRAL Spectrometer,''
  Astrophys.\ J.\  {\bf 646}, 965 (2006)
  [arXiv:astro-ph/0604277].
  %%CITATION = ASJOA,646,965;%%

%\cite{Churazov:2006bk}
\bibitem{Churazov:2006bk}
  E.~Churazov {\it et al.},
  %``INTEGRAL observations of the cosmic X-ray background in the 5-100 keV range
  %via occultation by the Earth,''
  Astron.\ Astrophys.\  {\bf 467}, 529 (2007) 
  arXiv:astro-ph/0608250.
  %%CITATION = ASTRO-PH/0608250;%%

%\cite{Weidenspointner:2000ab}
\bibitem{Weidenspointner:2000ab}
  G.~Weidenspointner {\it et al.},
  AIP Conf.\ Proc.\  {\bf 510}, 467 (2000).

%\cite{Sreekumar:1997un}
\bibitem{Sreekumar:1997un}
  P.~Sreekumar {\it et al.}  [EGRET Collaboration],
  %``EGRET observations of the extragalactic gamma ray emission,''
  Astrophys.\ J.\  {\bf 494}, 523 (1998)
  [arXiv:astro-ph/9709257].
  %%CITATION = ASJOA,494,523;%%
\cite{Strong:2004ry}

\bibitem{Strong:2004ry}
  A.~W.~Strong, I.~V.~Moskalenko and O.~Reimer,
  %``A new determination of the extragalactic diffuse gamma-ray background from
  %EGRET data,''
  Astrophys.\ J.\  {\bf 613}, 956 (2004)
  [arXiv:astro-ph/0405441].
  %%CITATION = ASJOA,613,956;%%





%\cite{Boyarsky:2008ju}
\bibitem{Boyarsky:2008ju}
  A.~Boyarsky, O.~Ruchayskiy and D.~Iakubovskyi,
  %``A lower bound on the mass of Dark Matter particles,''
  JCAP {\bf 0903}, 005 (2009)
  [arXiv:0808.3902 [hep-ph]].
  %%CITATION = JCAPA,0903,005;%%


%\cite{Kawasaki:1997ah}
\bibitem{Kawasaki:1997ah}
  M.~Kawasaki and T.~Yanagida,
  %``Constraint on cosmic density of the string moduli field in  gauge-mediated
  %supersymmetry-breaking theories,''
  Phys.\ Lett.\  B {\bf 399}, 45 (1997)
  [arXiv:hep-ph/9701346].
  %%CITATION = PHLTA,B399,45;%%

%\cite{Picciotto:2004rp}
\bibitem{Picciotto:2004rp}
  C.~Picciotto and M.~Pospelov,
  %``Unstable relics as a source of galactic positrons,''
  Phys.\ Lett.\  B {\bf 605}, 15 (2005)
  [arXiv:hep-ph/0402178].
  %%CITATION = PHLTA,B605,15;%%

%\cite{PalomaresRuiz:2007ry}
\bibitem{PalomaresRuiz:2007ry}
  S.~Palomares-Ruiz,
  %``Model-Independent Bound on the Dark Matter Lifetime,''
  Phys.\ Lett.\  B {\bf 665}, 50 (2008)
  [arXiv:0712.1937 [astro-ph]].
  %%CITATION = PHLTA,B665,50;%%

%\cite{Beacom:2005qv}
\bibitem{Beacom:2005qv}
  J.~F.~Beacom and H.~Yuksel,
  %``Stringent Constraint on Galactic Positron Production,''
  Phys.\ Rev.\ Lett.\  {\bf 97}, 071102 (2006)
  [arXiv:astro-ph/0512411].
  %%CITATION = PRLTA,97,071102;%%

%\cite{Sizun:2006uh}
\bibitem{Sizun:2006uh}
  P.~Sizun, M.~Casse and S.~Schanne,
  %``Continuum gamma-ray emission from light dark matter positrons and
  %electrons,''
  Phys.\ Rev.\  D {\bf 74}, 063514 (2006)
  [arXiv:astro-ph/0607374].
  %%CITATION = PHRVA,D74,063514;%%


%\cite{Britton:1992xv}
\bibitem{Britton:1992xv}
  D.~I.~Britton {\it et al.},
  %``Improved search for massive neutrinos in $\pi^+ \rightarrow e^+ \nu$
  %decay,''
  Phys.\ Rev.\  D {\bf 46}, 885 (1992).
  %%CITATION = PHRVA,D46,885;%%

%\cite{Hooper:2008im}
\bibitem{Hooper:2008im}
  D.~Hooper and K.~M.~Zurek,
  %``Natural supersymmetric model with MeV dark matter,''
  Phys.\ Rev.\  D {\bf 77}, 087302 (2008)
  [arXiv:0801.3686 [hep-ph]].
  %%CITATION = PHRVA,D77,087302;%%

%\cite{Boehm:2003ha}
\bibitem{Boehm:2003ha}
  C.~Boehm, P.~Fayet and J.~Silk,
  %``Light and heavy dark matter particles,''
  Phys.\ Rev.\  D {\bf 69}, 101302 (2004)
  [arXiv:hep-ph/0311143].
  %%CITATION = PHRVA,D69,101302;%%


%\cite{Boehm:2002yz}
\bibitem{Boehm:2002yz}
  C.~Boehm, T.~A.~Ensslin and J.~Silk,
  %``Are light annihilating dark matter particles possible?,''
  J.\ Phys.\ G {\bf 30}, 279 (2004)
  [arXiv:astro-ph/0208458].
  %%CITATION = JPHGB,G30,279;%%

%\cite{Boehm:2003hm}
\bibitem{Boehm:2003hm}
  C.~Boehm and P.~Fayet,
  %``Scalar dark matter candidates,''
  Nucl.\ Phys.\  B {\bf 683}, 219 (2004)
  [arXiv:hep-ph/0305261].
  %%CITATION = NUPHA,B683,219;%%

%\cite{Boehm:2003bt}
\bibitem{Boehm:2003bt}
  C.~Boehm, D.~Hooper, J.~Silk, M.~Casse and J.~Paul,
  %``MeV dark matter: Has it been detected?,''
  Phys.\ Rev.\ Lett.\  {\bf 92}, 101301 (2004)
  [arXiv:astro-ph/0309686].
  %%CITATION = PRLTA,92,101301;%%

%\cite{Knodlseder:2005yq}
\bibitem{Knodlseder:2005yq}
  J.~Knodlseder {\it et al.},
  %``The all-sky distribution of 511-keV electron positron annihilation
  %emission,''
  Astron.\ Astrophys.\  {\bf 441}, 513 (2005)
  [arXiv:astro-ph/0506026].
  %%CITATION = AAEJA,441,513;%%


%\cite{Jean:2005af}
\bibitem{Jean:2005af}
  P.~Jean {\it et al.},
  %``Spectral analysis of the Galactic e+e- annihilation emission,''
  Astron.\ Astrophys.\  {\bf 445}, 579 (2006)
  [arXiv:astro-ph/0509298].
  %%CITATION = AAEJA,445,579;%%


%\cite{Hooper:2003sh}
\bibitem{Hooper:2003sh}
  D.~Hooper, F.~Ferrer, C.~Boehm, J.~Silk, J.~Paul, N.~W.~Evans and M.~Casse,
  %``MeV dark matter in dwarf spheroidals: A smoking gun?,''
  Phys.\ Rev.\ Lett.\  {\bf 93}, 161302 (2004)
  [arXiv:astro-ph/0311150].
  %%CITATION = PRLTA,93,161302;%%

%\cite{Cordier:2004hf}
\bibitem{Cordier:2004hf}
  B.~Cordier {\it et al.},
  %``Search for a light dark matter annihilation signal in the Sagittarius
  %dwarf galaxy,''
  arXiv:astro-ph/0404499.
  %%CITATION = ASTRO-PH/0404499;%%




%\cite{Ascasibar:2005rw}
\bibitem{Ascasibar:2005rw}
  Y.~Ascasibar, P.~Jean, C.~Boehm and J.~Knoedlseder,
  %``Constraints on dark matter and the shape of the Milky Way dark halo from
  %the 511 keV line,''
  Mon.\ Not.\ Roy.\ Astron.\ Soc.\  {\bf 368}, 1695 (2006)
  [arXiv:astro-ph/0507142].
  %%CITATION = MNRAA,368,1695;%%

%\cite{Beacom:2004pe}
\bibitem{Beacom:2004pe}
  J.~F.~Beacom, N.~F.~Bell and G.~Bertone,
  %``Gamma-ray constraint on Galactic positron production by MeV dark  matter,''
  Phys.\ Rev.\ Lett.\  {\bf 94}, 171301 (2005)
  [arXiv:astro-ph/0409403].
  %%CITATION = PRLTA,94,171301;%%

%\cite{Boehm:2006df}
\bibitem{Boehm:2006df}
  C.~Boehm and P.~Uwer,
  %``Revisiting bremsstrahlung emission associated with light dark matter
  %annihilations,''
  arXiv:hep-ph/0606058.
  %%CITATION = HEP-PH/0606058;%%

%\cite{Lee:1977ua}
\bibitem{Lee:1977ua}
  B.~W.~Lee and S.~Weinberg,
  %``Cosmological lower bound on heavy-neutrino masses,''
  Phys.\ Rev.\ Lett.\  {\bf 39}, 165 (1977).
  %%CITATION = PRLTA,39,165;%%


%\cite{Fayet:2004bw}
\bibitem{Fayet:2004bw}
  P.~Fayet,
  %``Light spin-1/2 or spin-0 dark matter particles,''
  Phys.\ Rev.\  D {\bf 70}, 023514 (2004)
  [arXiv:hep-ph/0403226].
  %%CITATION = PHRVA,D70,023514;%%


%\cite{Rasera:2005sa}
\bibitem{Rasera:2005sa}
  Y.~Rasera, R.~Teyssier, P.~Sizun, B.~Cordier, J.~Paul, M.~Casse and P.~Fayet,
  %``Soft gamma-ray background and light dark matter annihilation,''
  Phys.\ Rev.\  D {\bf 73}, 103518 (2006)
  [arXiv:astro-ph/0507707].
  %%CITATION = PHRVA,D73,103518;%%


%\cite{Serpico:2004nm}
\bibitem{Serpico:2004nm}
  P.~D.~Serpico and G.~G.~Raffelt,
  %``MeV-mass dark matter and primordial nucleosynthesis,''
  Phys.\ Rev.\  D {\bf 70}, 043526 (2004)
  [arXiv:astro-ph/0403417].
  %%CITATION = PHRVA,D70,043526;%%

%\cite{Schleicher:2008gm}
\bibitem{Schleicher:2008gm}
  D.~R.~G.~Schleicher, S.~C.~O.~Glover, R.~Banerjee and R.~S.~Klessen,
  %``Cosmic constraints rule out s-wave annihilation of light dark matter,''
  Phys.\ Rev.\  D {\bf 79}, 023515 (2009)
  [arXiv:0809.1523 [astro-ph]].
  %%CITATION = PHRVA,D79,023515;%%

%\cite{Boehm:2004gt}
\bibitem{Boehm:2004gt}
  C.~Boehm and Y.~Ascasibar,
  %``More evidence in favour of light dark matter particles?,''
  Phys.\ Rev.\  D {\bf 70}, 115013 (2004)
  [arXiv:hep-ph/0408213].
  %%CITATION = PHRVA,D70,115013;%%


%\cite{Ahn:2005ck}
\bibitem{Ahn:2005ck}
  K.~Ahn and E.~Komatsu,
  %``Dark matter annihilation: The origin of cosmic gamma-ray background at
  %1-MeV to 20-MeV,''
  Phys.\ Rev.\  D {\bf 72}, 061301 (2005)
  [arXiv:astro-ph/0506520].
  %%CITATION = PHRVA,D72,061301;%%

%\cite{Bell:2008vx}
\bibitem{Bell:2008vx}
  N.~F.~Bell and T.~D.~Jacques,
  %``Gamma-ray Constraints on Dark Matter Annihilation into Charged Particles,''
  arXiv:0811.0821 [astro-ph].
  %%CITATION = ARXIV:0811.0821;%%

%\cite{Fayet:2007ua}
\bibitem{Fayet:2007ua}
  P.~Fayet,
  %``U-boson production in e+ e- annihilations, psi and Upsilon decays, and
  %light dark matter,''
  Phys.\ Rev.\  D {\bf 75}, 115017 (2007)
  [arXiv:hep-ph/0702176].
  %%CITATION = PHRVA,D75,115017;%%

%\cite{Fayet:2006xd}
\bibitem{Fayet:2006xd}
  P.~Fayet,
  %``U-boson detectability, and light dark matter,''
  arXiv:hep-ph/0607094.
  %%CITATION = HEP-PH/0607094;%%

%\cite{McElrath:2005bp}
\bibitem{McElrath:2005bp}
  B.~McElrath,
  %``Invisible quarkonium decays as a sensitive probe of dark matter,''
  Phys.\ Rev.\  D {\bf 72}, 103508 (2005)
  [arXiv:hep-ph/0506151].
  %%CITATION = PHRVA,D72,103508;%%

%\cite{Borodatchenkova:2005ct}
\bibitem{Borodatchenkova:2005ct}
  N.~Borodatchenkova, D.~Choudhury and M.~Drees,
  %``Probing MeV dark matter at low-energy e+ e- colliders,''
  Phys.\ Rev.\ Lett.\  {\bf 96}, 141802 (2006)
  [arXiv:hep-ph/0510147].
  %%CITATION = PRLTA,96,141802;%%

%\cite{Zhu:2007zt}
\bibitem{Zhu:2007zt}
  S.~h.~Zhu,
  %``U-boson at BESIII,''
  Phys.\ Rev.\  D {\bf 75}, 115004 (2007)
  [arXiv:hep-ph/0701001].
  %%CITATION = PHRVA,D75,115004;%%







%\cite{Fayet:2008cn}
\bibitem{Fayet:2008cn}
  P.~Fayet,
  %``U(1)_A symmetry in two-doublet models, U bosons or light pseudoscalars, and
  %psi and Upsilon decays,''
  arXiv:0812.3980 [hep-ph].
  %%CITATION = ARXIV:0812.3980;%%

%\cite{Fayet:2006sp}
\bibitem{Fayet:2006sp}
  P.~Fayet,
  %``Constraints on light dark matter and U bosons, from psi, Upsilon, K+, pi0,
  %eta and eta' decays,''
  Phys.\ Rev.\  D {\bf 74}, 054034 (2006)
  [arXiv:hep-ph/0607318].
  %%CITATION = PHRVA,D74,054034;%%

%\cite{Ablikim:2006eg}
\bibitem{Ablikim:2006eg}
  M.~Ablikim {\it et al.}  [BES Collaboration],
  %``Search for Invisible Decays of $\eta$ and $\eta^\prime$ in $J/\psi \to
  %\phi\eta$ and $\phi \eta^\prime$,''
  Phys.\ Rev.\ Lett.\  {\bf 97}, 202002 (2006)
  [arXiv:hep-ex/0607006].
  %%CITATION = PRLTA,97,202002;%%




%\cite{Boehm:2004uq}
\bibitem{Boehm:2004uq}
  C.~Boehm,
  %``Implications of a new light gauge boson for neutrino physics,''
  Phys.\ Rev.\  D {\bf 70}, 055007 (2004)
  [arXiv:hep-ph/0405240].
  %%CITATION = PHRVA,D70,055007;%%

%\cite{Boehm:2007na}
\bibitem{Boehm:2007na}
  C.~Boehm and J.~Silk,
  %``A new test of the light dark matter hypothesis,''
  Phys.\ Lett.\  B {\bf 661}, 287 (2008)
  [arXiv:0708.2768 [hep-ph]].
  %%CITATION = PHLTA,B661,287;%%


%\cite{Yuksel:2007ac}
\bibitem{Yuksel:2007ac}
  H.~Yuksel, S.~Horiuchi, J.~F.~Beacom and S.~Ando,
  %``Neutrino Constraints on the Dark Matter Total Annihilation Cross Section,''
  Phys.\ Rev.\  D {\bf 76}, 123506 (2007)
  [arXiv:0707.0196 [astro-ph]].
  %%CITATION = PHRVA,D76,123506;%%


%\cite{PalomaresRuiz:2007eu}
\bibitem{PalomaresRuiz:2007eu}
  S.~Palomares-Ruiz and S.~Pascoli,
  %``Testing MeV dark matter with neutrino detectors,''
  Phys.\ Rev.\  D {\bf 77}, 025025 (2008)
  [arXiv:0710.5420 [astro-ph]].
  %%CITATION = PHRVA,D77,025025;%%





%\cite{Barbieri:1988av}
\bibitem{Barbieri:1988av}
  R.~Barbieri and R.~N.~Mohapatra,
  %``LIMITS ON RIGHT-HANDED INTERACTIONS FROM SN1987A OBSERVATIONS,''
  Phys.\ Rev.\  D {\bf 39}, 1229 (1989).
  %%CITATION = PHRVA,D39,1229;%%



%\cite{Fayet:2006sa}
\bibitem{Fayet:2006sa}
  P.~Fayet, D.~Hooper and G.~Sigl,
  %``Constraints on light dark matter from core-collapse supernovae,''
  Phys.\ Rev.\ Lett.\  {\bf 96}, 211302 (2006)
  [arXiv:hep-ph/0602169].
  %%CITATION = PRLTA,96,211302;%%

%\cite{Hooper:2007tu}
\bibitem{Hooper:2007tu}
  D.~Hooper, M.~Kaplinghat, L.~E.~Strigari and K.~M.~Zurek,
  %``MeV Dark Matter and Small Scale Structure,''
  Phys.\ Rev.\  D {\bf 76}, 103515 (2007)
  [arXiv:0704.2558 [astro-ph]].
  %%CITATION = PHRVA,D76,103515;%%

%\cite{Caldwell:1981rj}
\bibitem{Caldwell:1981rj}
  J.~A.~R.~Caldwell and J.~P.~Ostriker,
  %``The Mass distribution within our Galaxy: A Three component model,''
  Astrophys.\ J.\  {\bf 251}, 61 (1981).
  %%CITATION = ASJOA,251,61;%%

%\cite{Kamionkowski:1997xg}
\bibitem{Kamionkowski:1997xg}
  M.~Kamionkowski and A.~Kinkhabwala,
  %``Galactic halo models and particle dark matter detection,''
  Phys.\ Rev.\  D {\bf 57}, 3256 (1998)
  [arXiv:hep-ph/9710337].
  %%CITATION = PHRVA,D57,3256;%%

%\cite{Strumia:2003zx}
\bibitem{Strumia:2003zx}
  A.~Strumia and F.~Vissani,
  %``Precise quasielastic neutrino nucleon cross section,''
  Phys.\ Lett.\  B {\bf 564}, 42 (2003)
  [arXiv:astro-ph/0302055].
  %%CITATION = PHLTA,B564,42;%%



%\cite{Frere:2006hp}
\bibitem{Frere:2006hp}
  J.~M.~Frere, F.~S.~Ling, L.~Lopez Honorez, E.~Nezri, Q.~Swillens and G.~Vertongen,
  %``MeV right-handed neutrinos and dark matter,''
  Phys.\ Rev.\  D {\bf 75}, 085017 (2007)
  [arXiv:hep-ph/0610240].
  %%CITATION = PHRVA,D75,085017;%%

%\cite{Abdurashitov:2000va}
\bibitem{Abdurashitov:2000va}
  J.~N.~Abdurashitov {\it et al.}  [SAGE Collaboration],
  %``Solar neutrino results from SAGE,''
  Phys.\ Atom.\ Nucl.\  {\bf 63}, 943 (2000)
  [Yad.\ Fiz.\  {\bf 63}, 1019 (2000)].
  %%CITATION = YAFIA,63,1019;%%

%\cite{Hampel:1998xg}
\bibitem{Hampel:1998xg}
  W.~Hampel {\it et al.}  [GALLEX Collaboration],
  %``GALLEX solar neutrino observations: Results for GALLEX IV,''
  Phys.\ Lett.\  B {\bf 447}, 127 (1999).
  %%CITATION = PHLTA,B447,127;%%



%\cite{Cleveland:1998nv}
\bibitem{Cleveland:1998nv}
  B.~T.~Cleveland {\it et al.},
  %``Measurement of the solar electron neutrino flux with the Homestake
  %chlorine detector,''
  Astrophys.\ J.\  {\bf 496}, 505 (1998).
  %%CITATION = ASJOA,496,505;%%

%\cite{Gando:2002ub}
\bibitem{Gando:2002ub}
  Y.~Gando {\it et al.}  [Super-Kamiokande Collaboration],
  %``Search for anti-electron-neutrinos from the Sun at Super-Kamiokande-I,''
  Phys.\ Rev.\ Lett.\  {\bf 90}, 171302 (2003)
  [arXiv:hep-ex/0212067].
  %%CITATION = PRLTA,90,171302;%%



%\cite{Eguchi:2003gg}
\bibitem{Eguchi:2003gg}
  K.~Eguchi {\it et al.}  [KamLAND Collaboration],
  %``A high sensitivity search for anti-nu/e's from the sun and other  sources
  %at KamLAND,''
  Phys.\ Rev.\ Lett.\  {\bf 92}, 071301 (2004)
  [arXiv:hep-ex/0310047].
  %%CITATION = PRLTA,92,071301;%%



%\cite{Balata:2006db}
\bibitem{Balata:2006db}
  M.~Balata {\it et al.}  [Borexino Collaboration],
  %``Search for electron antineutrino interactions with the Borexino  counting
  %test facility at Gran Sasso,''
  Eur.\ Phys.\ J.\  C {\bf 47}, 21 (2006)
  [arXiv:hep-ex/0602027].
  %%CITATION = EPHJA,C47,21;%%




%\cite{Aharmim:2006wq}
\bibitem{Aharmim:2006wq}
  B.~Aharmim {\it et al.}  [SNO Collaboration],
  %``A search for neutrinos from the solar hep reaction and the diffuse
  %supernova neutrino background with the sudbury neutrino observatory,''
  Astrophys.\ J.\  {\bf 653}, 1545 (2006)
  [arXiv:hep-ex/0607010].
  %%CITATION = ASJOA,653,1545;%%

%\cite{Lunardini:2008xd}
\bibitem{Lunardini:2008xd}
  C.~Lunardini and O.~L.~G.~Peres,
  %``Upper limits on the diffuse supernova neutrino flux from the
  %SuperKamiokande data,''
  JCAP {\bf 0808}, 033 (2008)
  [arXiv:0805.4225 [astro-ph]].
  %%CITATION = JCAPA,0808,033;%%


%\cite{Raby:2008pd}
\bibitem{Raby:2008pd}
  S.~Raby {\it et al.},
  %``DUSEL Theory White Paper,''
  arXiv:0810.4551 [hep-ph].
  %%CITATION = ARXIV:0810.4551;%%

%\cite{Learned:2008zj}
\bibitem{Learned:2008zj}
  J.~G.~Learned, S.~T.~Dye and S.~Pakvasa,
  %``Hanohano: A Deep Ocean Anti-Neutrino Detector for Unique Neutrino Physics
  %and Geophysics Studies,''
  arXiv:0810.4975 [hep-ex].
  %%CITATION = ARXIV:0810.4975;%%







\end{thebibliography}

\end{document}